# Bright nonblinking photoluminescence with blinking lifetime from a nanocavity-coupled quantum dot

Zhiyuan Wang[1], Jianwei Tang[1,2]*, Jiahao Han[1], Juan Xia[1], Tianzi Ma[1],

Xue-Wen Chen[1,2]*

[1]School of Physics, Wuhan National Laboratory for Optoelectronics, Institute for Quantum Science and Engineering and Hubei Key Laboratory of Gravitation and Quantum Physics, Huazhong University of Science and Technology, Wuhan 430074, P. R. China

[2]Wuhan Institute of Quantum Technology, Wuhan 430206, P. R. China

*Corresponding authors:

J. T. (jianwei_tang@hust.edu.cn), X.-W. C. (xuewen_chen@hust.edu.cn)

**Abstract:**

Colloidal semiconductor quantum dots (QDs) are excellent luminescent nanomaterials for a broad range of optoelectronic applications. Their photoluminescence blinking, however, hinders their practical use in many aspects. It has been shown that coupling QDs to plasmonic nanostructures may provide a viable way to suppress blinking. Nevertheless, the underlying mechanism of blinking suppression remains unclear and debated. Here, by deterministically coupling a single QD to a plasmonic nanocavity, we clarify the mechanism of blinking suppression, and demonstrate unprecedentedly bright emission from a single colloidal QD. In particular, we report for the first time that the coupled system exhibits nonblinking photoluminescence with blinking lifetime, which shows that the elimination of photoluminescence blinking originates from enhanced quantum yield of the charged states. We identify that the radiative decay rate is boosted from $(48 \text{ ns})^{-1}$ to $(0.7 \text{ ns})^{-1}$, which outcompetes Auger processes and enables similar quantum yields for charged and neutral excitons. Moreover, we demonstrate ultrabright photoluminescence of up to 17 million detected photons per second from a single QD. This work sheds new light on the goal of achieving ultrabright nonblinking QDs and may benefit a variety of QD-based applications.

# 1. Introduction

Colloidal quantum dots (QDs) represent a class of semiconductor nanocrystals exhibiting unique quantum confinement effects[1-3]. As quantum emitters, they exhibit remarkable characteristics, including large absorption cross sections[4,5], near-unity quantum yields[5-7], enhanced photostability compared to dye molecules[8], and tunable wavelengths[9]. Such exceptional properties make QDs highly promising for various fields, as evidenced by their impact on lighting, displays, imaging, lasers, energy, and information[10]. However, QDs exhibit stochastic photoluminescence (PL) intermittency or "blinking"[11], detrimental for many applications.

Since the discovery of QD blinking in 1996[12], tremendous efforts have aimed to understand the mechanisms and suppress it[13]. Proposed QD blinking mechanisms remain debated, but a widely accepted model considers it a result of stochastic switching between neutral and charged states[14]. In order to suppress QD blinking, one may decrease the blinking frequency by reducing the charging probability, and/or decrease the blinking amplitude by increasing the charged state quantum yields. Charging probability can be reduced through surface passivation to eliminate surface charge traps[15,16] or through growth of high-quality shell layers with high confinement potentials to separate excited charge carriers from surface traps[17]. Charged state quantum yields can be increased by suppressing Auger processes, which can be accomplished by constructing a weak[18-22] or gradient[23-25] confinement for charge carriers.

Despite the impressive progress in blinking suppression through chemical engineering, such methods impose certain limitations on QD materials, structures, and synthesis. Consequently, researchers have explored optical approaches for suppressing QD blinking. In 2002, Bawendi et al. discovered that coupling QDs to rough metal surfaces could enhance radiative decay rates enough to outcompete Auger relaxation, restoring the quantum yield of charged excitons close to that of neutral excitons[26]. Since then, coupling to various plasmonic nanostructures has enabled significant blinking suppression for diverse QDs[26-37].

However, the mechanism for plasmon-enabled blinking suppression remains unclear, with debates on whether the suppression stems from the enhancement of charged state quantum yield[27-31] or suppression of QD charging[32-34] or both[26,35-37]. Understanding the underlying mechanism is crucial to devise cost-effective strategies to suppress QD blinking in various occasions. To elucidate the mechanism, the experiments should ideally meet the following conditions: (1) unambiguous same-QD comparison, (2) strong Purcell effect through single-mode coupling, (3) purely optical effects when coupled to plasmonic nanostructures. However, most studies investigating blinking suppression lack same-QD comparison, which hampers quantitative analyses due to the substantial variability between QDs. The few reports that have employed same-QD comparisons utilized gold nanospheres as plasmonic nanostructures[29,32], which required the QD-metal distances to be extremely small in order to achieve strong Purcell effects. This causes significant coupling into nonradiative high-order multipolar modes[38], as well as introduces chemical and charge transfer effects[39], complicating the interpretation.

The coupling to plasmonic nanostructures provides the additional advantage of enhanced radiative decay rates[40,41]. This enhancement holds the potential for achieving high brightness, provided that the coupled system exhibits a high radiation efficiency and can withstand strong excitation. However, previous demonstrations of sub-nanosecond radiative decay, aimed at suppressing blinking, often fail to ensure dominant coupling to the radiative dipolar mode, resulting in pronounced plasmonic quenching [26,29,32]. Furthermore, the photochemical stability of QDs is generally insufficient to sustain operation under strong excitation[42]. QDs typically undergo photobleaching well before reaching saturation excitation. Due to plasmonic quenching and/or photobleaching, the highest reported brightness for a single colloidal quantum dot (QD) is currently limited to the level of 1 million detected photons per second[26,29,31,42,43].

In this study, by coupling a silica-encapsulated QD to the dipolar mode of a plasmonic nanocavity, we unambiguously demonstrate blinking suppression, elucidate the underlying mechanism, and report single-QD PL brightness of 17 million detected photons per second. The QD's lifetime is shortened from 48 ns to 0.7 ns due to the Purcell effect. This is sufficient for the radiative decay to outcompete Auger relaxation and thereby boosts the quantum yield of charged excitons close to neutral excitons, enabling nonblinking PL. Crucially, we observe lifetime blinking in the nanocavity-enabled nonblinking PL. The lifetime occasionally dropped by half. The binary fluctuation of lifetime indicates continued charging and discharging, and therefore provides compelling evidence that the mechanism for the removal of PL blinking is enhancement of charged state quantum yields rather than complete inhibition of charging. In the experiment, in-situ coupling via nanomanipulation facilitates before-after comparison on the same QD. Effects that can complicate interpretation are minimized by single-mode nanocavity-QD coupling and silica encapsulation of QDs. These enable us to establish a quantitative agreement between experiment and the proposed mechanism.

## 2. Results and Discussion

### 2.1. Deterministic single-mode nanocavity-QD coupling

The proposed design of the nanocavity-QD coupled system is illustrated in **Figure 1**a. A single QD encapsulated in silica (TEM image in Figure 1a; Note S1, Supporting Information) is positioned at the hot spot of a nanocavity composed of a dimer of gold nanorods (AuNRs; TEM image in Figure 1a). This design enables a strong Purcell effect through single-mode coupling to the dipolar mode of the nanocavity[44]. The silica shell surrounding the QD serves as a passivating layer, enhancing the QD's photostability and minimizing chemical and charge transfer effects during nanocavity-QD coupling. The nanocavity-QD coupled system can be deterministically constructed by assembling nanoparticles through nanomanipulation using an atomic force microscope (AFM)[29,45,46]. Subsequently, optical characterizations can be carried out in situ using the experimental setup depicted in Figure 1a, which integrates an AFM with an inverted optical microscope (Note S2, Supporting Information). This facilitates unambiguous same-QD comparison before and after nanocavity-QD coupling.

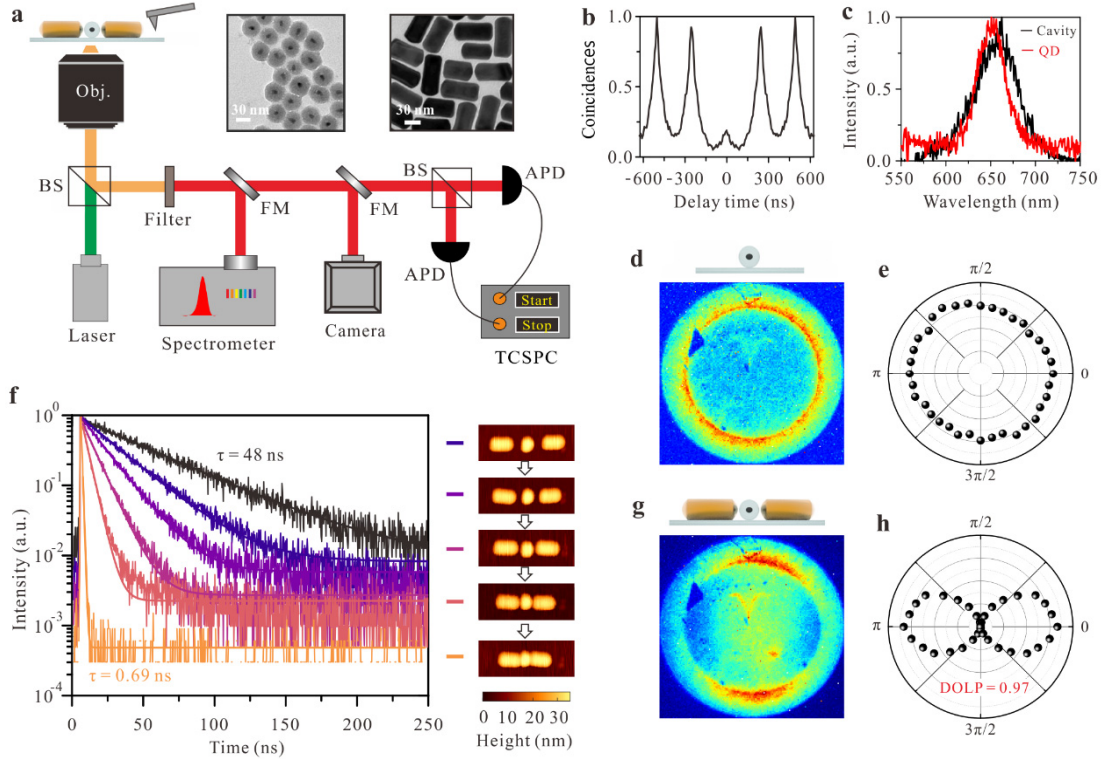

**Figure 1. Deterministic single-mode nanocavity-QD coupling. a**) Experimental setup for deterministic QD-nanocavity coupling and optical characterization. Insets: TEM images of silica-encapsulated QDs (left) and AuNRs (right). BS beam splitter, FM flip mirror, APD avalanche photodiode single photon detector, TCSPC time-correlated single photon counting. **b**) Second-order correlation function $g^{(2)}(\tau)$ of the PL photons from the selected QD. **c**) PL spectrum of the QD (red) and dark-field scattering spectrum of the nanocavity (black). **d**) Far-field emission pattern (measured at the back-focal plane of the objective) of the QD before coupling with the nanocavity. **e**) Far-field emission polarization state measurement for the QD before coupling with the nanocavity. **f**) PL decay curves for the QD before (black) and after (color-coded for different coupling stages shown by the AFM topographic images in the inset) coupling with the nanocavity. **g**) Far-field emission pattern (measured at the back-focal plane of the objective) of the QD coupled with the nanocavity. **h**) Far-field emission polarization state measurement for the QD coupled with the nanocavity, representing a high degree of linear polarization (DOLP) of 0.97.

Initially, AuNRs and QDs were dispersed on a glass coverslip by spin casting their colloidal solutions. A single QD was then selected, and its single-QD behavior was confirmed by the observation of anti-bunching in its second-order correlation function (Figure 1b). The emission spectrum of the QD displayed a central wavelength of 650 nm and a linewidth of 20 nm (Figure 1c). The QD's far-field emission pattern, measured at the back focal plane of the objective, exhibited a nearly isotropic ring shape (Figure 1d). Consistent with this emission pattern, the polarization state measurement indicated near-unpolarized emission (Figure 1e), suggesting that the QD's 2D dipole lay horizontally[47]. The QD's lifetime was determined to be 48 ns by fitting the PL decay curve (black curve in Figure 1f).

Subsequently, two AuNRs were selected and manipulated using the AFM tip to approach the QD, thereby constructing the hybrid system (Note S3, Supporting Information). The gap size of the nanocavity was tuned stepwise, as shown by the AFM topographic images in Figure 1f. As the gap width decreased, the decay rate increased progressively, as indicated by the color-coded curves in Figure 1f. When the gap size was minimized, the QD's lifetime was finally shortened from 48 ns to 0.69 ns, corresponding to a Purcell factor of 70. This Purcell effect is strong enough for the radiative decay to outcompete the Auger relaxations. The dipolar resonance of the nanocavity aligns with the emission spectrum of the QD, as shown by the dark-field scattering spectrum depicted in Figure 1c. Numerical simulation indicates that the strong Purcell effect is contributed almost exclusively by the single-mode coupling to the nanocavity's dipolar mode (Note S4, Supporting Information). Consequently, the measured emission pattern (Figure 1g) follows that of a linear dipole, and the polarization measurement (Figure 1h) shows a high degree of linear polarization (DOLP) of 0.97.

## 2.2. Nanocavity-enabled nonblinking PL with blinking lifetime

Next, we investigate the nanocavity's impact on the QD's blinking behavior by comparing the correlated intensity-lifetime trajectories of the same QD before and after its coupling with the nanocavity (**Figure 2**).

Prior to coupling with the nanocavity, the QD exhibited typical blinking in its PL intensity trajectory (Figure 2a), accompanied by simultaneous fluctuations in the lifetime trajectory (Figure 2b). Analysis of the intensity distribution (right panel of Figure 2a) and lifetime distribution (right panel of Figure 2b) clearly reveals three distinct brightness states and three lifetime states. The lifetime-intensity distribution[48] plotted in Figure 2c exhibits three spots, which indicates that the brightness states and lifetime states are correlated and can be attributed to three charging states, the neutral state (denoted as "X") and the positively and negatively charged states (denoted as $X^{*A}$ and $X^{*B}$). The charged states have lower quantum yields and faster decay rates than the neutral state. Based on the intensity distribution, we estimate the quantum yields of the charged states as $\eta^{*A} = 0.33\eta$ and $\eta^{*B} = 0.17\eta$, where $\eta$ represents the quantum yield at the neutral state. The PL decay curves for these states, displayed in Figure 2d, all exhibit monoexponential decays with rates $\gamma_{tot} = (48\ ns)^{-1}$, $\gamma_{tot}^{*A} = (7.8\ ns)^{-1}$, and $\gamma_{tot}^{*B} = (4.1\ ns)^{-1}$, respectively.

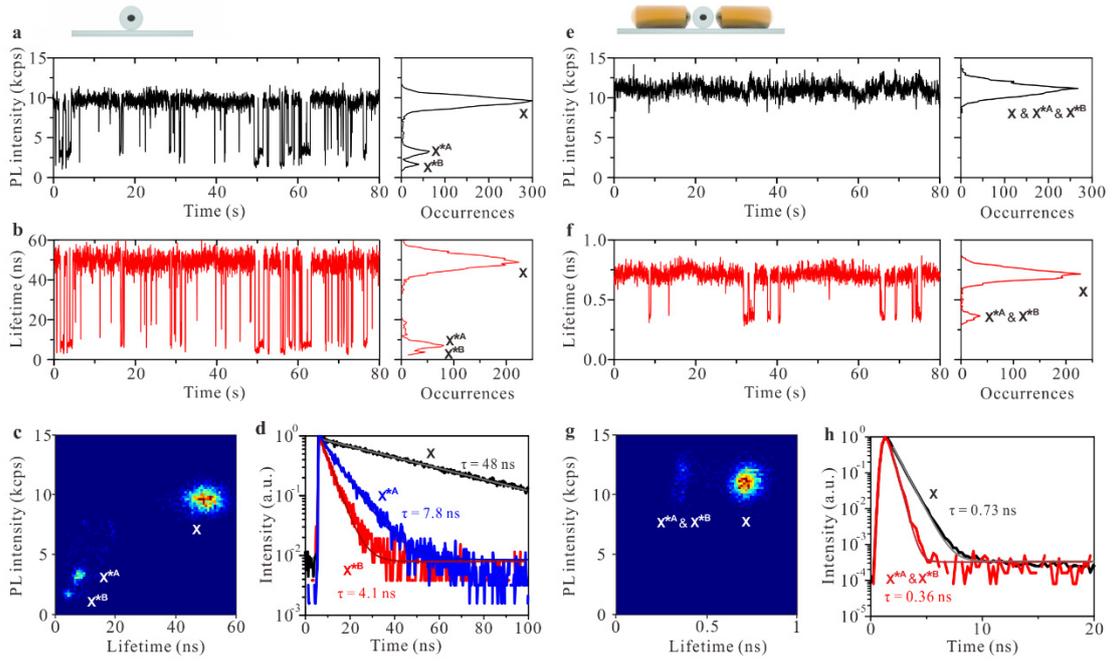

**Figure 2. Nanocavity-enabled nonblinking PL with blinking lifetime. a**, **b**) PL intensity trajectory (a) and PL lifetime trajectory (b) of the single QD before coupling with the nanocavity. A statistical distribution is plotted to the right of each trajectory. **c**) PL lifetime-intensity distribution of the QD before coupling with the nanocavity. **d**) PL decay curves at the neutral state (black) and the charged states (blue and red) of the QD before coupling with the nanocavity. **e**, **f**, **g** and **h** are the corresponding results after coupling the QD with the nanocavity.

Upon coupling with the nanocavity, the radiative decay rate was remarkably enhanced to a sub-nanosecond level (Figure 1f). The enhanced radiative decay can outcompete Auger relaxations and thereby keep the quantum yield high even at the charged states. This shall lead to suppression of blinking. A nonblinking intensity trajectory and the corresponding lifetime trajectory are displayed in Figure 2e and 2f, respectively. Intriguingly, the nonblinking intensity trajectory is accompanied by binary fluctuations in lifetime. The lifetime occasionally drops by half. This behavior is visually represented as two spots in the lifetime-intensity distribution graph depicted in Figure 2g, corresponding to two states with similar intensity but distinct lifetimes. The PL decay curves for the two states are displayed in Figure 2h, both displaying monoexponential decays. The slower state exhibits a decay rate of $(0.73\ \text{ns})^{-1}$, while the faster state has a doubled decay rate of $(0.36\ \text{ns})^{-1}$.

To the best of our knowledge, this is the first demonstration of nanocavity-enabled nonblinking single-QD PL with lifetime blinking. The observed fluctuations in lifetime indicate transitions between different charging states. We attribute the slower state to the neutral state and the faster state to the charged states. The lifetime fluctuation provides clear evidence that the absence of blinking in the PL intensity trajectory is not due to complete inhibition of charging, but rather a result of improved quantum yields of the charged states, as will be further confirmed through quantitative analysis. Nevertheless, it should be noted that this does not exclude the possibility of charging suppression. When comparing Figure 2f to Figure 2b, the diminished frequency of lifetime blinking does indicate a decrease in charging probability, albeit not a complete inhibition. This suppression of charging may be explained by the Purcell-enhanced decay rate[35]. The enhanced decay rate reduces the average time the electron/hole spends in the higher energy state, thereby decreasing the probability of charge transfer to the traps.

**2.3. Quantitative agreement between experiment and theory**

In the following, we establish a quantitative agreement between the experimental observations and the theoretical modelling, corroborating the proposed mechanism for nanocavity-enabled nonblinking PL. We first quantitatively determine, based on the correlated intensity-lifetime analysis before nanocavity coupling (Figure 2a-2d), the impact of an additional electron or hole on the relaxation rates, including the doubling of the radiative decay rate and the opening of the Auger relaxation channels.

The presence of an additional charge carrier, either an electron or a hole, in the charged states doubles the number of radiative recombination pathways. In the ideal case of statistical scaling, this would result in doubled radiative decay rates[49,50]. However, Coulombic interactions between charge carriers may cause deviation from doubling[51]. To account for this, we express the enhanced radiative decay rates as $\gamma_r^{*A} = \beta^{*A}\gamma_r$ and $\gamma_r^{*B} = \beta^{*B}\gamma_r$, where $\gamma_r$ represents the radiative decay rate at the neutral state and the enhancement factors $\beta^{*A}$ and $\beta^{*B}$ are yet to be determined. Considering the quantum yields of the neutral and charged states as $\eta = \gamma_r/\gamma_{tot}$, $\eta^{*A} = \beta^{*A}\gamma_r/\gamma_{tot}^{*A}$, and $\eta^{*B} = \beta^{*B}\gamma_r/\gamma_{tot}^{*B}$, we calculate $\beta^{*A} = (\eta^{*A}/\eta)(\gamma_{tot}^{*A}/\gamma_{tot}) = 2.03$ and $\beta^{*B} = (\eta^{*B}/\eta)(\gamma_{tot}^{*B}/\gamma_{tot}) = 1.99$. These values are very close to the theoretical value of 2.[50,51]

The added electron or hole also opens up Auger relaxation pathway, with rates denoted as $\gamma_A^{*A}$ and $\gamma_A^{*B}$. Assuming a unity quantum yield for the neutral state ($\eta = 1$)[5-7], the total decay rates of the neutral and charged states can be expressed as $\gamma_{tot} = \gamma_r$, $\gamma_{tot}^{*A} = \beta^{*A}\gamma_r + \gamma_A^{*A}$, and $\gamma_{tot}^{*B} = \beta^{*B}\gamma_r + \gamma_A^{*B}$, respectively. With the measured total decay rates (Figure 2d), we obtain $\gamma_r = \gamma_{tot} = (48 \text{ ns})^{-1}$, $\gamma_A^{*A} = \gamma_{tot}^{*A} - \beta^{*A}\gamma_r = (11.6 \text{ ns})^{-1}$, and $\gamma_A^{*B} = \gamma_{tot}^{*B} - \beta^{*B}\gamma_r = (4.9 \text{ ns})^{-1}$. The distinct values of $\gamma_A^{*A}$ and $\gamma_A^{*B}$ arise from differences in localization and density of states between holes and electrons[51].

With the Auger rates determined and the radiative rate doubling effect confirmed, we then quantitatively check the blinking suppression mechanism with the correlated intensity-lifetime analysis after nanocavity coupling (Figure 2e-2h). Assuming a unity quantum yield for the neutral state[5-7], we can express the decay rates of the neutral and charged states as $\gamma_{tot,cav} = F\gamma_r$, $\gamma_{tot,cav}^{*A} = \beta^{*A}F\gamma_r + \gamma_A^{*A}$ and $\gamma_{tot,cav}^{*B} = \beta^{*B}F\gamma_r + \gamma_A^{*B}$, respectively, where $F$ represents the Purcell factor. Based on the measured neutral state decay rate $F\gamma_r = (0.73 \text{ ns})^{-1}$, we determine the decay rates of the charged states as $\gamma_{tot,cav}^{*A} = (0.35 \text{ ns})^{-1}$ and $\gamma_{tot,cav}^{*B} = (0.34 \text{ ns})^{-1}$, aligning with the measured decay rate $(0.36 \text{ ns})^{-1}$ for the faster state. The two charged states, which originally had significantly different decay rates due to disparate Auger rates (Figure 2b-2d), now possess nearly equal decay rates, indistinguishable in the lifetime analysis (Figure 2f-2h). The impact of the added charge is now primarily manifested in the doubling of the radiative decay rate, while the Auger pathway opened by the added charge is negligible. Accordingly, charging induces minimal quantum yield reduction. The quantum yields of the charged states can be calculated as $\eta_{\text{cav}}^{*A} = \beta^{*A}F\gamma_r/(\beta^{*A}F\gamma_r + \gamma_A^{*A}) = 0.97$ and $\eta_{\text{cav}}^{*B} = \beta^{*B}F\gamma_r/(\beta^{*B}F\gamma_r + \gamma_A^{*B}) = 0.93$, respectively. They are indeed very close to unity, to the extent that they cannot be distinguished from the neutral state in the intensity trajectory (Figure 2e).

**2.4. Bright single-QD PL**

The PL brightness of colloidal QDs is generally limited by their long intrinsic PL lifetime, typically in the range of tens of nanoseconds. Since we have achieved blinking suppression by significantly increasing the radiative transition rate to sub-nanosecond levels, the issue of long lifetime has been addressed in the process. Nonetheless, attaining high brightness requires fulfilling additional conditions. First, it is crucial to predominantly couple the QD to the highly radiative dipolar mode of the nanocavity, avoiding coupling to nonradiative multipolar modes that cause plasmonic quenching of PL. In addition, a high photochemical stability is required for QDs to withstand intense laser irradiation. A previous report showed that encapsulating QDs with a gold shell can enhance their photochemical stability for operation under strong excitation[30]. However, the lifetime remained above 10 ns due to the mild Purcell effect the gold shell can provide, yielding a maximum single-QD PL brightness of only 2 million detected photons per second.

In our study, the strong Purcell effect enabling sub-nanosecond lifetime was achieved without compromising single-mode coupling to the dipolar mode, thus minimizing plasmonic quenching. Furthermore, blinking suppression and silica encapsulation enhanced QD's photochemical stability. These factors fulfilled the necessary conditions for ultrabright single-QD PL. **Figure 3**a illustrates the detected PL intensity of the nanocavity-coupled single QD as a function of the excitation power. As the excitation power increases, the PL intensity initially exhibits linear growth, gradually entering the saturation region, ultimately reaching an intensity of 17 million detected photons per second. During the measurement, the QD was exposed only for a short duration under each excitation power. The intensity trajectories corresponding to the data points in Figure 3a are displayed in Figure 3b. In order to achieve long-lasting high-brightness emission, it is imperative to make further advancements in the stability of QDs. This entails improving QD's structure and crystal quality[21,52], as well as optimizing the environment where QDs are situated, such as embedding the hybrid structure in a polymer film for enhanced encapsulation[17,53].

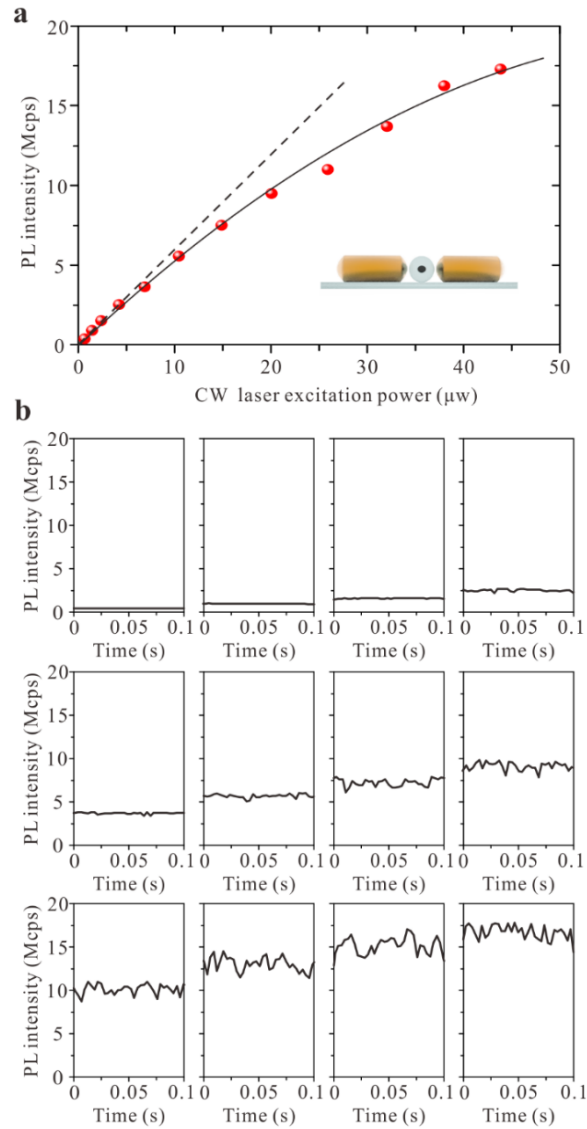

**Figure 3. Bright single-QD PL. a**) Detected PL intensity of the nanocavity-coupled single QD as a function of the excitation power. The solid line represents a fit of the data points, while the dashed line is a linear extrapolation given for a visual guide. **b**) PL intensity trajectories corresponding to the data points in (a). The power-dependent background signals are subtracted.

## 3. Conclusion

In conclusion, we have experimentally demonstrated that coupling a QD to the dipolar mode of a plasmonic dimer nanocavity can substantially boost the radiative decay to outcompete the Auger relaxation, achieving similar quantum yields for charged and neutral monoexcitons, and consequently enabling nonblinking PL. Notably, the observation of lifetime blinking in the nanocavity-enabled nonblinking PL provides direct evidence against the plausible mechanism of complete inhibition of charging, and supports the mechanism of enhanced quantum yields of charged states. The mechanism is further corroborated by the quantitative agreement between theoretical calculations and experimental observations. The clarification of the mechanism contributes to the fundamental understanding of QD blinking, and may have practical implications for the development of highly efficient and stable QD-based devices.

Furthermore, we have demonstrated the great potential for enhancing single-QD brightness by reducing lifetime while maintaining high radiation efficiency and adequate photostability under intense excitation. The achieved 17 MHz photon detection rate implies a photon generation rate approaching the gigahertz level, considering the collection and detection efficiency of approximately 3%. Bright, nonblinking single QDs hold tremendous potential for a wide range of QD-based applications. Additionally, the demonstrated linear dipolar emission with high DOLP is also highly desirable for applications. It can enable deterministic dipole orientation control for efficient mode coupling when integrating the coupled QD-nanocavity systems into advanced photonic platforms, such as on-chip photonic circuits[54,55] and fiber-based networks[31,56]. This represents a promising avenue for future exploration. It is also crucial to dedicate more efforts to enhancing the photochemical stability of QDs. For large-scale fabrication of the coupled QD-nanocavity systems, alternative approaches such as lithography[43,57,58] and self-assembly[59,60] can be considered.

## Supporting Information

Supporting Information is available from the Wiley Online Library or from the author.

## Acknowledgements

We gratefully acknowledge financial support from the National Natural Science Foundation of China (Grant Number 92150111, 62235006, 12004130, 12374349, 62105112), the Science and Technology Department of Hubei Province, China (Project No. 2022BAA018), the China Postdoctoral Science Foundation (2021M700048) and Huazhong University of Science and Technology.

## Conflict of Interest

The authors declare no conflict of interest.

## Data Availability Statement

The data that support the findings of this study are available from the corresponding authors upon reasonable request.

## References

1. Ekimov, A. I. Quantum size effect in three-dimensional microscopic semiconductor crystals. *JETP Lett.* **34**, 345 (1981).
2. Rossetti, R., Nakahara, S. & Brus, L. E. Quantum size effects in the redox potentials, resonance Raman spectra, and electronic spectra of CdS crystallites in aqueous solution. *The Journal of Chemical Physics* **79**, 1086-1088 (1983).
3. Murray, C. B., Norris, D. J. & Bawendi, M. G. Synthesis and characterization of nearly monodisperse CdE (E = sulfur, selenium, tellurium) semiconductor nanocrystallites. *Journal of the American Chemical Society* **115**, 8706-8715 (1993).
4. Leatherdale, C. A., Woo, W. K., Mikulec, F. V. & Bawendi, M. G. On the Absorption Cross Section of CdSe Nanocrystal Quantum Dots. *The Journal of Physical Chemistry B* **106**, 7619-7622 (2002).
5. Xu, W. W., Hou, X. Q., Meng, Y. J., Meng, R. Y., Wang, Z. Y., Qin, H. Y., Peng, X. G. & Chen, X. W. Deciphering Charging Status, Absolute Quantum Efficiency, and Absorption Cross Section of Multicarrier States in Single Colloidal Quantum Dots. *Nano Letters* **17**, 7487-7493 (2017).
6. Brokmann, X., Coolen, L., Dahan, M. & Hermier, J. P. Measurement of the radiative and nonradiative decay rates of single CdSe nanocrystals through a controlled modification of their spontaneous emission. *Physical Review Letters* **93**, 107403 (2004).
7. Spinicelli, P., Buil, S., Quélin, X., Mahler, B., Dubertret, B. & Hermier, J. P. Bright and Grey States in CdSe-CdS Nanocrystals Exhibiting Strongly Reduced Blinking. *Physical Review Letters* **102**, 136801 (2009).
8. van Sark, W., Frederix, P., Bol, A. A., Gerritsen, H. C. & Meijerink, A. Blueing, bleaching, and blinking of single CdSe/ZnS quantum dots. *Chemphyschem* **3**, 871-879 (2002).
9. Smith, A. M. & Nie, S. M. Semiconductor Nanocrystals: Structure, Properties, and Band Gap Engineering. *Accounts of Chemical Research* **43**, 190-200 (2010).
10. de Arquer, F. P. G., Talapin, D. V., Klimov, V. I., Arakawa, Y., Bayer, M. & Sargent, E. H.

Semiconductor quantum dots: Technological progress and future challenges. *Science* **373**, 640 (2021).

11. Frantsuzov, P., Kuno, M., Janko, B. & Marcus, R. A. Universal emission intermittency in quantum dots, nanorods and nanowires. *Nature Physics* **4**, 519-522 (2008).

12. Nirmal, M., Dabbousi, B. O., Bawendi, M. G., Macklin, J. J., Trautman, J. K., Harris, T. D. & Brus, L. E. Fluorescence intermittency in single cadmium selenide nanocrystals. *Nature* **383**, 802-804 (1996).

13. Efros, A. L. & Nesbitt, D. J. Origin and control of blinking in quantum dots. *Nature Nanotechnology* **11**, 661-671 (2016).

14. Cordones, A. A. & Leone, S. R. Mechanisms for charge trapping in single semiconductor nanocrystals probed by fluorescence blinking. *Chemical Society Reviews* **42**, 3209-3221 (2013).

15. Hohng, S. & Ha, T. Near-complete suppression of quantum dot blinking in ambient conditions. *Journal of the American Chemical Society* **126**, 1324-1325 (2004).

16. Hammer, N. I., Early, K. T., Sill, K., Odoi, M. Y., Emrick, T. & Barnes, M. D. Coverage-mediated suppression of blinking in solid state quantum dot conjugated organic composite nanostructures. *Journal of Physical Chemistry B* **110**, 14167-14171 (2006).

17. Cao, H. J., Ma, J. L., Huang, L., Qin, H. Y., Meng, R. Y., Li, Y. & Peng, X. G. Design and Synthesis of Antiblinking and Antibleaching Quantum Dots in Multiple Colors via Wave Function Confinement. *Journal of the American Chemical Society* **138**, 15727-15735 (2016).

18. Chen, Y., Vela, J., Htoon, H., Casson, J. L., Werder, D. J., Bussian, D. A., Klimov, V. I. & Hollingsworth, J. A. "Giant" multishell CdSe nanocrystal quantum dots with suppressed blinking. *Journal of the American Chemical Society* **130**, 5026-5027 (2008).

19. Mahler, B., Spinicelli, P., Buil, S., Quelin, X., Hermier, J. P. & Dubertret, B. Towards non-blinking colloidal quantum dots. *Nature Materials* **7**, 659-664 (2008).

20. Ghosh, Y., Mangum, B. D., Casson, J. L., Williams, D. J., Htoon, H. & Hollingsworth, J. A. New Insights into the Complexities of Shell Growth and the Strong Influence of Particle Volume in Nonblinking "Giant" Core/Shell Nanocrystal Quantum Dots. *Journal of the American Chemical Society* **134**, 9634-9643 (2012).

21. Chen, O., Zhao, J., Chauhan, V. P., Cui, J., Wong, C., Harris, D. K., Wei, H., Han, H. S., Fukumura, D., Jain, R. K. & Bawendi, M. G. Compact high-quality CdSe-CdS core-shell nanocrystals with narrow emission linewidths and suppressed blinking. *Nature Materials* **12**, 445-451 (2013).

22. Qin, H. Y., Niu, Y., Meng, R. Y., Lin, X., Lai, R. C., Fang, W. & Peng, X. G. Single-Dot Spectroscopy of Zinc-Blende CdSe/CdS Core/Shell Nanocrystals: Nonblinking and Correlation with Ensemble Measurements. *Journal of the American Chemical Society* **136**, 179-187 (2014).

23. Cragg, G. E. & Efros, A. L. Suppression of Auger Processes in Confined Structures. *Nano Letters* **10**, 313-317 (2010).

24. Park, Y. S., Bae, W. K., Padilha, L. A., Pietryga, J. M. & Klimov, V. I. Effect of the Core/Shell Interface on Auger Recombination Evaluated by Single-Quantum-Dot Spectroscopy. *Nano Letters* **14**, 396-402 (2014).

25. Nasilowski, M., Spinicelli, P., Patriarche, G. & Dubertret, B. Gradient CdSe/CdS Quantum Dots with Room Temperature Biexciton Unity Quantum Yield. *Nano Letters* **15**, 3953-3958 (2015).

26. Shimizu, K. T., Woo, W. K., Fisher, B. R., Eisler, H. J. & Bawendi, M. G. Surface-enhanced emission from single semiconductor nanocrystals. *Physical Review Letters* **89**, 117401 (2002).

27. Fu, Y., Zhang, J. & Lakowicz, J. R. Suppressed blinking in single quantum dots (QDs) immobilized near silver island films (SIFs). *Chemical Physics Letters* **447**, 96-100 (2007).

28. Yuan, C. T., Yu, P., Ko, H. C., Huang, J. & Tang, J. Antibunching Single-Photon Emission and Blinking Suppression of CdSe/ZnS Quantum Dots. *ACS Nano* **3**, 3051-3056 (2009).
29. Ratchford, D., Shafiei, F., Kim, S., Gray, S. K. & Li, X. Q. Manipulating Coupling between a Single Semiconductor Quantum Dot and Single Gold Nanoparticle. *Nano Letters* **11**, 1049-1054 (2011).
30. Ji, B. T., Giovanelli, E., Habert, B., Spinicelli, P., Nasilowski, M., Xu, X. Z., Lequeux, N., Hugonin, J. P., Marquier, F., Greffet, J. J. & Dubertret, B. Non-blinking quantum dot with a plasmonic nanoshell resonator. *Nature Nanotechnology* **10**, 170-175 (2015).
31. Shafi, K. M., Yalla, R. & Nayak, K. P. Bright and Polarized Fiber In-Line Single-Photon Source Based on Plasmon-Enhanced Emission into Nanofiber Guided Modes. *Physical Review Applied* **19**, 034008 (2023).
32. Bharadwaj, P. & Novotny, L. Robustness of Quantum Dot Power-Law Blinking. *Nano Letters* **11**, 2137-2141 (2011).
33. Lu, L., Tong, X., Zhang, X., Ren, N. F., Jiang, B. & Lu, H. F. Hot spot assisted blinking suppression of CdSe quantum dots. *Chemical Physics Letters* **652**, 167-171 (2016).
34. Al Masud, A., Arefin, S. M. N., Fairooz, F., Fu, X., Moonschi, F., Srijanto, B. R., Neupane, K. R., Aryal, S., Calabro, R., Kim, D. Y., Collier, C. P., Chowdhury, M. H. & Richards, C. I. Photoluminescence Enhancement, Blinking Suppression, and Improved Biexciton Quantum Yield of Single Quantum Dots in Zero Mode Waveguides. *Journal of Physical Chemistry Letters* **12**, 3303-3311 (2021).
35. Ma, X. D., Tan, H., Kipp, T. & Mews, A. Fluorescence Enhancement, Blinking Suppression, and Gray States of Individual Semiconductor Nanocrystals Close to Gold Nanoparticles. *Nano Letters* **10**, 4166-4174 (2010).
36. Jiang, Q. B., Roy, P., Claude, J. B. & Wenger, J. Single Photon Source from a Nanoantenna-Trapped Single Quantum Dot. *Nano Letters* **21**, 7030-7036 (2021).
37. He, Y. L., Chen, J., Liu, R. M., Weng, Y. L., Zhang, C., Kuang, Y. M., Wang, X. J., Guo, L. J. & Ran, X. Suppressed Blinking and Polarization-Dependent Emission Enhancement of Single ZnCdSe/ZnS Dot Coupled with Au Nanorods. *ACS Applied Materials & Interfaces* **14**, 12901-12910 (2022).
38. Delga, A., Feist, J., Bravo-Abad, J. & Garcia-Vidal, F. J. Quantum Emitters Near a Metal Nanoparticle: Strong Coupling and Quenching. *Physical Review Letters* **112**, 253601 (2014).
39. Park, Y. S., Ghosh, Y., Chen, Y., Piryatinski, A., Xu, P., Mack, N. H., Wang, H. L., Klimov, V. I., Hollingsworth, J. A. & Htoon, H. Super-Poissonian Statistics of Photon Emission from Single CdSe-CdS Core-Shell Nanocrystals Coupled to Metal Nanostructures. *Physical Review Letters* **110**, 117401 (2013).
40. Wei, H., Yan, X. H., Niu, Y. J., Li, Q., Jia, Z. L. & Xu, H. X. Plasmon-Exciton Interactions: Spontaneous Emission and Strong Coupling. *Advanced Functional Materials* **31**, 2100889 (2021).
41. Qian, Z., Shan, L., Zhang, X., Liu, Q., Ma, Y., Gong, Q. & Gu, Y. Spontaneous emission in micro- or nanophotonic structures. *PhotoniX* **2**, 21 (2021).
42. Hoang, T. B., Akselrod, G. M. & Mikkelsen, M. H. Ultrafast Room-Temperature Single Photon Emission from Quantum Dots Coupled to Plasmonic Nanocavities. *Nano Letters* **16**, 270-275 (2016).
43. Urena, E. B., Kreuzer, M. P., Itzhakov, S., Rigneault, H., Quidant, R., Oron, D. & Wenger, J. Excitation Enhancement of a Quantum Dot Coupled to a Plasmonic Antenna. *Advanced Materials* **24**, OP314-OP320 (2012).
44. Meng, Y., Huang, D., Li, H., Feng, X., Li, F., Liang, Q., Ma, T., Han, J., Tang, J., Chen, G. & Chen, X.-W. Bright single-nanocrystal upconversion at sub 0.5 W cm−2 irradiance via coupling to single nanocavity mode. *Nature Photonics* **17**, 73-81 (2023).

45. Tang, J. W., Xia, J., Fang, M. D., Bao, F. L., Cao, G. J., Shen, J. Q., Evans, J. & He, S. L. Selective far-field addressing of coupled quantum dots in a plasmonic nanocavity. *Nature Communications* **9**, 1705 (2018).
46. Xia, J., Tang, J., Bao, F., Sun, Y., Fang, M., Cao, G., Evans, J. & He, S. Turning a hot spot into a cold spot: polarization-controlled Fano-shaped local-field responses probed by a quantum dot. *Light: Science & Applications* **9**, 166 (2020).
47. Empedocles, S. A., Neuhauser, R. & Bawendi, M. G. Three-dimensional orientation measurements of symmetric single chromophores using polarization microscopy. *Nature* **399**, 126-130 (1999).
48. Galland, C., Ghosh, Y., Steinbruck, A., Sykora, M., Hollingsworth, J. A., Klimov, V. I. & Htoon, H. Two types of luminescence blinking revealed by spectroelectrochemistry of single quantum dots. *Nature* **479**, 203 (2011).
49. McGuire, J. A., Joo, J., Pietryga, J. M., Schaller, R. D. & Klimov, V. I. New Aspects of Carrier Multiplication in Semiconductor Nanocrystals. *Accounts of Chemical Research* **41**, 1810-1819 (2008).
50. Galland, C., Ghosh, Y., Steinbruck, A., Hollingsworth, J. A., Htoon, H. & Klimov, V. I. Lifetime blinking in nonblinking nanocrystal quantum dots. *Nature Communications* **3**, 908 (2012).
51. Park, Y. S., Bae, W. K., Pietryga, J. M. & Klimov, V. I. Auger Recombination of Biexcitons and Negative and Positive Trions in Individual Quantum Dots. *ACS Nano* **8**, 7288-7296 (2014).
52. Qin, H. Y., Meng, R. Y., Wang, N. & Peng, X. G. Photoluminescence Intermittency and Photo-Bleaching of Single Colloidal Quantum Dot. *Advanced Materials* **29**, 1606923 (2017).
53. Schlegel, G., Bohnenberger, J., Potapova, I. & Mews, A. Fluorescence decay time of single semiconductor nanocrystals. *Physical Review Letters* **88**, 137401 (2002).
54. Elsinger, L., Gourgues, R., Zadeh, I. E., Maes, J., Guardiani, A., Bulgarini, G., Pereira, S. F., Dorenbos, S. N., Zwiller, V., Hens, Z. & Van Thourhout, D. Integration of Colloidal PbS/CdS Quantum Dots with Plasmonic Antennas and Superconducting Detectors on a Silicon Nitride Photonic Platform. *Nano Letters* **19**, 5452-5458 (2019).
55. Shlesinger, I., Palstra, I. M. & Koenderink, A. F. Integrated Sideband-Resolved SERS with a Dimer on a Nanobeam Hybrid. *Physical Review Letters* **130**, 016901 (2023).
56. Sugawara, M., Xuan, Y. N., Mitsumori, Y., Edamatsu, K. & Sadgrove, M. Plasmon-enhanced single photon source directly coupled to an optical fiber. *Physical Review Research* **4**, 043146 (2022).
57. Curto, A. G., Volpe, G., Taminiau, T. H., Kreuzer, M. P., Quidant, R. & van Hulst, N. F. Unidirectional Emission of a Quantum Dot Coupled to a Nanoantenna. *Science* **329**, 930-933 (2010).
58. Santhosh, K., Bitton, O., Chuntonov, L. & Haran, G. Vacuum Rabi splitting in a plasmonic cavity at the single quantum emitter limit. *Nature Communications* **7**, 11823 (2016).
59. Ko, S. H., Du, K. & Liddle, J. A. Quantum-Dot Fluorescence Lifetime Engineering with DNA Origami Constructs. *Angewandte Chemie-International Edition* **52**, 1193-1197 (2013).
60. Nicoli, F., Zhang, T., Hübner, K., Jin, B., Selbach, F., Acuna, G., Argyropoulos, C., Liedl, T. & Pilo-Pais, M. DNA-Mediated Self-Assembly of Plasmonic Antennas with a Single Quantum Dot in the Hot Spot. *Small* **15**, 1804418 (2019).